\documentclass[twocolumn,superscriptaddress,showpacs,amsmath,amssymb,pra,longbibliography]{revtex4-1}

\usepackage[pdftex]{graphicx} 
\usepackage{dcolumn}  
\usepackage{bm}       
\usepackage[usenames,dvipsnames]{color}
\definecolor{URLCOL}{rgb}{0,0.52,0.83} 
\definecolor{LINKCOL}{rgb}{0.05,0.5,0} 
\definecolor{orange}{rgb}{0.6,0.3,0} 
\definecolor{CITECOL}{rgb}{0.25,0,0.48} 
\usepackage{epstopdf}

\usepackage[normalem]{ulem}

\usepackage{booktabs}
\usepackage{natbib}
\usepackage{tabularx}
\usepackage{rotating}

\def\en{_{\rm en}}

\def\unpol{^{\rm unpol}}
\def\pol{^{\rm pol}}
\def\w{z}
\def\n{\nu}

\def\H{_\mathrm{H}}

\graphicspath{{f/}}

\definecolor{TITLECOL}{rgb}{0.1,0.2,0.7} 
\definecolor{SECOL}{rgb}{0.1,0.2,0.7} 
\definecolor{CONTENTSCOL}{rgb}{0.1,0.2,0.7} 
\definecolor{SSECOL}{rgb}{0.25,0,0.48} 
\definecolor{SSSECOL}{rgb}{0.2,0.08,0.53} 
\definecolor{FINCOL}{rgb}{0.01,0.3,0.07} 

\definecolor{URLCOL}{rgb}{0,0.17,0.43} 
\definecolor{LINKCOL}{rgb}{0.05,0.4,0} 
\definecolor{CITECOL}{rgb}{0.35,0,0.48} 

\def\sss{\scriptscriptstyle\rm}

\def\bea{\begin{eqnarray}}
\def\eea{\end{eqnarray}}
\def\ben{\begin{equation}}
\def\een{\end{equation}}
\def\benu{\begin{enumerate}}
\def\enu{\end{enumerate}}

\def\bei{\begin{itemize}}
\def\eei{\end{itemize}}
\def\beit{\begin{itemize}}
\def\eit{\end{itemize}}
\def\benu{\begin{enumerate}}
\def\enu{\end{enumerate}}

\def\x{_{\sss X}}
\def\c{_{\sss C}}
\def\s{_{\sss S}}

\def\xc{_{\sss XC}}

\def\H{_{\sss H}}
\def\ee{_{\rm ee}}

\def\LDA{^{\rm LDA}}

\def\unif{^{\rm unif}}
\def\up{_\uparrow}
\def\dn{_\downarrow}

\def\e{_\text{e}}

\begin{document}

\title{One Dimensional Mimicking of Electronic Structure: The Case for Exponentials}
\author{Thomas E.~Baker}
\affiliation{Department of Physics \& Astronomy, University of California, Irvine, California 92697 USA}
\author{E.~Miles Stoudenmire}
\affiliation{Perimeter Institute for Theoretical Physics, Waterloo, Ontario, N2L 2Y5 Canada}
\author{Lucas O.~Wagner}
\affiliation{Department of Theoretical Chemistry and Amsterdam Center for Multiscale Modeling,FEW, Vrije Universiteit, De Boelelaan 1083, 1081HV Amsterdam, The Netherlands}
\author{Kieron Burke}
\affiliation{Department of Chemistry, University of California, Irvine, California 92697 USA}
\affiliation{Department of Physics \& Astronomy, University of California, Irvine, California 92697 USA}
\author{Steven R.~White}
\affiliation{Department of Physics \& Astronomy, University of California, Irvine, California 92697 USA}
\date{\today}

\begin{abstract}
An exponential interaction is constructed so that one-dimensional atoms and chains of atoms mimic the general behavior of their three-dimensional counterparts.  Relative to the more commonly used soft-Coulomb interaction, the exponential greatly diminishes the computational time needed for calculating highly accurate quantities with the density matrix renormalization group.  This is due to the use of a small matrix product operator and to exponentially vanishing tails.  Furthermore, its more rapid decay closely mimics the screened Coulomb interaction in three dimensions.  Choosing parameters to best match earlier calculations, we report results for the one dimensional hydrogen atom, uniform gas, and small atoms and molecules both exactly and in the local density approximation.

 \end{abstract}


\maketitle

\section{Introduction}\label{intro}

The notion that a strong electromagnetic field can be modeled well by a one dimensional (1d) problem dates back to at least 1939 when a calculation by Schiff and Snyder \cite{SS39} assigned a potential in the $x$ direction and ``integrated out" the transverse degrees of freedom in the $y$ and $z$ directions by averaging over radial wave functions.  The resulting interaction has been studied by several others \cite{EL60,KG92,KG94,SCSM08,SCSM09,CSS06} including one work by Elliott and Loudon \cite{EL60} that approximated the 1d potential, so it can be solved more easily.

Today, the most common approximation for the 1d potentials in a strong electromagnetic field was introduced by Eberly, Su, and Javaninen \cite{ESJ89} who rewrote the Coulomb interaction $1/\sqrt{x^2+y^2+z^2}$ as $1/\sqrt{x^2+a^2}$; the $x$ component is allowed to vary independently and the remaining radial term in cylindrical coordinates, $a$, is set to a constant which is determined on a system-by-system basis for the particular application of study to match the ionization energy. This soft-Coulomb interaction has been used in a wide variety of applications requiring a 1d potential \cite{ESJ89,LSE91,joachain2012atoms,KLEG01,LGE00,CCZB96,CC89,B93book,BW92}.  It has many attractive features including the avoidance of the singularity at zero separation while retaining a Rydberg series of excitations \cite{NSS11,GW14,CNSS15}. Even when not considering a strong electromagnetic field, the soft-Coulomb interaction has become the choice potential for 1d model systems.

A general use of 1d systems is as a computational laboratory for studying the limitations of electronic structure methods, such as density functional theory (DFT) \cite{JGP93,JGP94,JGGP94,KG94,UG94,CLSO03,MB04,CBKK07,VFCC08,LG09,HFCV11,BW12,WSBW12,SWWB12,WSBW13,WBSB14}.  A key requirement is that the accuracy of such methods be at least qualitatively similar to their three dimensional (3d) counterparts. Helbig, {\it et.~al.} \cite{HFCV11} had already calculated the energy of the uniform electron gas with a soft-Coulomb repulsion, making the construction of the local density approximation (LDA) simple. Recently, some of us \cite{WSBW12} used Ref.~\onlinecite{HFCV11} to construct benchmark systems with the soft-Coulomb interaction which we used to study DFT approximations \cite{HK64,KS65,B12,BW13}. It was also shown that the model 1d systems closely mimic 3d systems, particularly those with high symmetry. The 1d nature of these systems allows us to use the density matrix renormalization group (DMRG) \cite{W92} to obtain ground states to numerical precision. DMRG has no sign problem and works well even for strongly correlated 1d systems \cite{S05}. We also apply various DFT approximations to these systems in order to understand how such approximations could be improved, especially when correlations are strong. We have found that our 1d ``pseudomolecules'' closely mimic the behavior of real 3d molecules in terms of the relative size and type of correlations and also the errors made by DFT approximations \cite{WSBW12}. Conclusions can be drawn from the 1d systems that are relevant to realistic 3d systems, a key advantage over lattice based models. Crucially, in 1d, we can extrapolate to the thermodynamic limit with far less computational cost compared to a similar calculation in 3d \cite{SW11}.

However, the soft-Coulomb interaction has some drawbacks as a mimic for 3d electronic structure calculations. First, the long $1/|x|$ tail has a bigger effect in 1d than in 3d, making the interaction excessively long ranged.   Second, in 3d, the electron-electron interaction induces weak cusps as $\mathbf{r}\rightarrow0$; the soft-Coulomb induces no cusps at all.  While this can be a significant computational advantage with some methods, this precludes using it to study cusp behavior.  Third, although the extra cost for treating power-law decaying interactions within DMRG can be made to scale sub-linearly with system size (via a clever 
fitting approximation \cite{M07,S11}), this cost is still much higher than for a system with strictly local interactions, for example, the Hubbard model \cite{Crosswhite:2008}.

As an alternative and complement to the soft-Coulomb interaction, we suggest an exponential interaction that addresses each of these weaknesses but whose parameters are chosen to give very similar results.  Its tails are weaker, simulating the shorter ranged nature of the Coulomb interaction in 3d if screening is taken into account. The presence of a cusp in the potential more accurately reflects 3d calculations, since a discontinuity in the wavefunction or its derivatives may make convergence more difficult. Finally, the cost for using exponential interactions within DMRG is no more than for using local interactions; in practice this makes DMRG calculations with 
exponential interactions more than an order of magnitude faster than with soft-Coulomb.
Extrapolations to the thermodynamic and continuum limits with exponential interactions also require less computational time.

The primary purpose of this paper is therefore to show in what ways an exponential interaction is preferable to the soft-Coulomb interaction in 1d electronic structure calculations and to provide reference results for such systems. Showing that the soft-Coulomb interaction is well approximated by the exponential would allow for the efficient and fast use of DMRG over QMC methods for 1d systems, so we construct our exponential to also mimic 1d soft-Coulomb calculations and allow for comparison with previous results \cite{L63,H81,H83,H67,HSC07,GG90,GC09,KFE98,T98,balentsnotes,S90,S93,AXW04,EG95,VFCC08,LSOC03}.

We derive some general analytic forms for the exponential hydrogenic atom in Sec.~\ref{expsec} and discuss mimicking the soft-Coulomb interaction with $a=1$ in Sec.~\ref{expsC}. Section~\ref{methods} summarizes the numerical tools used to find accurate results of the exponential systems. We find the uniform gas energies with DMRG and derive high and low density asymptotic limits in Sec.~\ref{Interact}.  Finite system calculations and benchmarks for the soft-Coulomb-like exponential interaction are given in Sec.~\ref{results}.  The Supplemental Material contains all necessary raw data to reproduce results \cite{BSWB15}.

\section{Hydrogenic Atoms}\label{expsec}

\subsection{General Analytic Expressions}\label{NonInt}

We wish to solve the 1d exponential Hydrogenic atom, with the potential:
\ben\label{ven}
v\en(x) = - Zv_\mathrm{exp}(x),\quad v_\mathrm{exp}(x)=A\exp(-\kappa|x|).
\een
where $A$ and $\kappa$ characterize the interaction and $Z$ is the `charge' felt by an electron (e) from a nucleus (n). Since the potential is even in $x$, the eigenstates
will be even and odd, and we label them $i=0,1,2...$ (alternating even to odd).  Writing
\ben
\w(x)=(2\sqrt{2AZ}/\kappa)\,\exp(-\kappa x/2),\quad x\geq0,
\een
the Schr\"odinger equation becomes the Bessel equation.  The wavefunction solution is a linear combination of $J_{\pm \n}(\w)$,
where $\n^2 = -8 E/\kappa^2$, and $J$ is a Bessel function of the first
kind.  The negative index ($-\nu$) functions diverge as $x\to\infty$, so the wavefunction is proportional to $J_{\n}(\w)$ and the density is
\ben
n(x)=C^2J^2_{\n}(\w(|x|))
\een
where $C$ is chosen so that $\int dx\,n(x)=1$.

The eigenvalues, $E$, are determined by spatial symmetry, so that:
\ben
\left.\frac d{dx}J_{\n}(\w)\right|_{\w_0}=0~~(\mathrm{even}), ~~~~~~J_{\n}(\w_0)=0~~(\mathrm{odd}),
\een
with $z_0=z(0)$. This condition implies that the eigenfunctions are Bessel functions of non-integer index. 
Defining $j'_i(\n)$ and $j_i(\n)$ to be zeroes of these
functions, indexed in order (see e.g., Ref. \cite{AS72}, Sec 9.5), the energy eigenvalues are
\ben
E_i = - \kappa^2\, \n^2_i(\w_0)/8,~~~~i=0,1,2,..
\een
where $\n_i(\w_0)$ satisfies
\ben
j'_{i+1}(\n_{2i}) = \w_0,~~~~
j_i(\n_{2i-1}) = \w_0.
\een
We found no source where these are generally listed
or approximated but they are available in, e.g., Mathematica.
One can deduce the critical values of $\w_0$
at which the number of bound states changes (when $E=0$), since $j'_1(0)=0$, $j'_2(0)=3.83171$,
$j'_3(0)=7.01559$, and $j_1(0)=2.40483$ and $j_2(0)=5.52008$
(Table 9.5 of Ref. \cite{AS72}).

Unlike a soft-Coulomb potential,
cusps appear for the exponential in analogy with 3d Coulomb systems. 
Rewriting the Schr\"odinger equation as $\psi''/(2\psi)=v\en(x)+E$, where $\psi$ is the
ground state wavefunction and primes denote derivatives, shows this ratio contains the cusp
of the potential
at the origin.

\subsection{Mimicking Soft-Coulomb Interactions}\label{expsC}

\begin{figure}
\includegraphics[width=1.1\columnwidth]{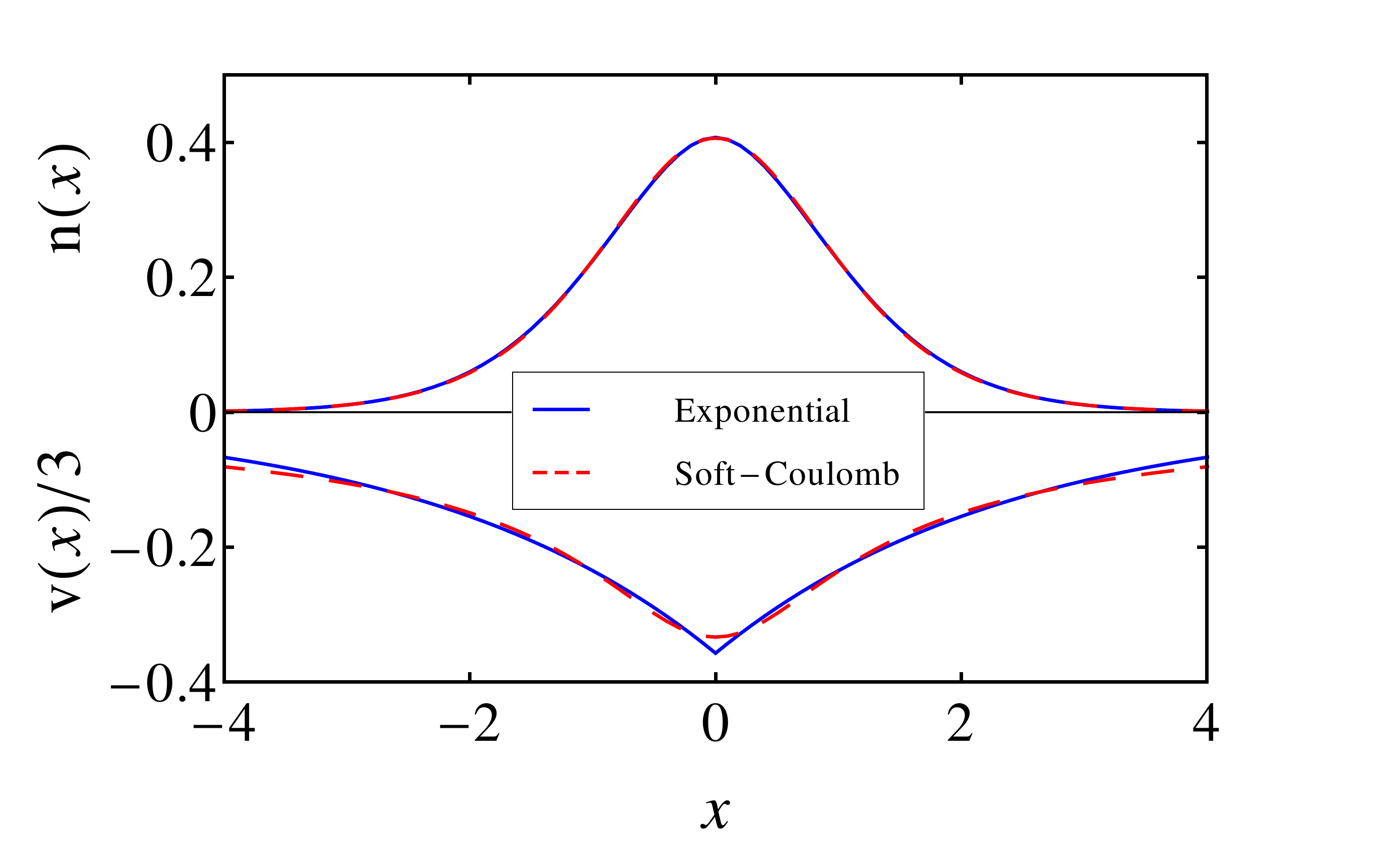}
\caption{(color online) An exponential interaction with 
$A=1.071295$ and $\kappa^{-1}=2.385345$ mimics the 
soft-Coulomb ($a=1$) hydrogen ground-state density very closely. \label{expLengthScale}}
\end{figure}

Now we choose the parameters of our exponential to
match closely those of the soft-Coulomb we have previously used 
in these studies, which has softness parameter $a=1$.  In that
case, the ground-state energy of the 1d H atom is 
-0.669778 to $\mu$Ha accuracy \cite{ESJ89,WSBW12,expEnote}, and 
the width of the ground state density, $\int dx\, n(x) x^2$, is 
1.191612.  We find $A$ and $\kappa$ of $v_\mathrm{exp}(x)$ to match these
values, yielding:
\begin{equation}\label{factors}
A=1.071295\quad\mathrm{and}\quad\kappa^{-1}=2.385345.
\end{equation}
Figure~\ref{expLengthScale} shows the close agreement between the
soft-Coulomb and the exponential in both the potential and the density for the
1d H atom.  Hereafter, we
take these values as defining our choice of exponential interaction. 

These values
for our hydrogen
atom yield $\w_0=6.983117$, so that it binds exactly four states.
Since $\w_0$ is proportional to ${\sqrt{Z}}$, where $Z$ is the nuclear
charge for a hydrogenic atom, a fifth state is just barely bound
when $Z = 1.00931$. This is a marked difference from the non-interacting soft-Coulomb \cite{L59} and non-interacting 3d hydrogen \cite{simon} atoms which each bind an infinite number of states.  

We then choose the repulsion between electrons
to be the same but with opposite sign:
\begin{equation}\label{expint}
v\ee(x-x')=v_{\mathrm{exp}}(x-x').
\end{equation}
Solutions for more than one electron are remarkably similar to the soft-Coulomb under these conditions.  For example, both the soft-Coulomb and the exponential bind neutral atoms only up to $Z=4$.  
We further note that H$^-$ is bound, but H$^{--}$ is not,
for all three cases (exponential, soft-Coulomb,
and reality).

\section{Calculational Details and Notation}\label{methods}

DMRG is an exceedingly efficient method for calculating low-energy properties of 1d systems and gives variational energies.
For the types of systems considered here, DMRG is able to determine essentially the exact ground state wavefunction \cite{S05,S11}. DMRG works by iteratively determining the minimal basis of many-body states needed to represent the ground state wavefunction to a given accuracy.
For 1d systems, one can reach excellent accuracy by retaining only a few hundred states in the basis.
The basis states are gradually improved by projecting the Hamiltonian into the current basis on all but a few lattice sites, 
then computing the ground state of this partially projected Hamiltonian. 
The new trial ground state is typically closer to the exact one, making it possible to compute an improved basis. 

Our calculations are done in real space on a grid \cite{itensor}.  The grid Hamiltonian is chosen to be the extended Hubbard-like model also used in Ref.~\onlinecite{WSBW12} (see Eq.~(3) of that work).  The electron-electron interaction and external potentials are changed to the exponentials used here. Note that the convention we use to evaluate integrals is to multiply the value of the integrand at each grid point by the grid spacing to match Ref.~\onlinecite{WSBW12}.  So, for here,
\begin{equation}
\int dx\,f(x)\approx\Delta\sum_if(x_i)
\end{equation}
for some function $f(x)$ and grid spacing $\Delta$.  Grid points are indexed by $i$ and span the interval of integration.

Using an exponential interaction within DMRG is very natural since it can be exactly represented by a matrix product operator,
the form of the Hamiltonian used in newer DMRG codes \cite{M07}. In fact, the soft-Coulomb interaction used in Ref.~\onlinecite{WSBW12} was actually represented by fitting it to approximately 25 exponentials \cite{PMCV10}. 

The majority of our focus is on finite systems and we use a finite grid with end points chosen
to make the wavefunction negligible.  For chains, we place the ends of the grid at
the first missing nucleus,
although in the case of a single atom or isolated molecule, we will place the
boundary sufficiently far from the nuclei to avoid noticeable effects.  This can
be much nearer to the nuclei than in the soft-Coulomb systems. 
We always use the convention of placing a grid point on a cusp when possible,
to avoid missing any kinetic energy.

In Kohn-Sham (KS) DFT \cite{KS65}, we use density functionals which are defined on the KS system, a non-interacting system that has the same density and energy as the full interacting system. Although, this non-interacting system has a different potential, $v_s(x)$, the KS potential.  The ground state energy, $E$, of an interacting electron system is the sum of the kinetic energy, $T$, the electron-electron repulsion, $V\ee$, and the one-body potential energy, $V$.  The ground state energy, $E$, is the minimization in $n$ of the density functional
\ben\label{E[n]}
E[n]=T\s[n]+U[n]+V[n]+E\xc[n],
\een
where
\ben
T\s[n]=\frac12\sum_\sigma\sum^{N_\sigma}_{i=1}\int dx\,\left|\frac d{dx}\phi_{i,\sigma}(x)\right|^2
\een
is the non-interacting kinetic energy evaluated on the occupied non-interacting KS orbitals, $\phi_{i,\sigma}$, for $N_\sigma$ particles of spin $\sigma$.  The difference, $T\c=T-T\s$, gives the difference between the interacting and non-interacting kinetic energy.  The Hartree integral is defined as
\ben
U[n]=\frac12\iint dx\,dx'\,v\ee(x-x')n(x)n(x').
\een
The one-body potential functional is
\ben
V[n]=\int dx\,v(x)n(x),
\een
where $v(x)$ is the attraction to the nuclei.  Equation~\ref{E[n]} defines the exchange-correlation (XC) functional, $E\xc[n]$, which is the sum of the exchange (X) and correlation (C) energies.  The exchange energy can be evaluated over the occupied KS orbitals as
\ben
\begin{split}
E\x[n]&=-\frac12\sum_\sigma\sum^{N_\sigma}_{i,j=1}\int dx \int dx'\,v\ee(x-x')\\
&\times\phi^*_{i,\sigma}(x)\phi^*_{j,\sigma}(x)\phi_{i,\sigma}(x')\phi_{j,\sigma}(x'),
\end{split}
\een
but in practice it is often approximted by an explicity density functional to reduce the computational cost.

\section{Uniform Gas}\label{Interact}

In order to construct and employ the LDA for our exponential
repulsion, we must first calculate the exchange-correlation energy of the exponentially
repelling uniform gas, expellium.

\subsection{Kinetic, Hartree, and Exchange Energies}\label{tshxc}

These energy components are straightforward and well known, as the single particle states are simply plane waves. The unpolarized, non-interacting kinetic energy
per length of a uniform gas is given by \cite{T27,F28,LS77}
\begin{equation}
t\unpol\s(n)=\pi^2n^3/24.
\end{equation}
The Hartree energy per length of a uniform system with exponential interaction is finite:
\ben
e\H\unif(n) = A\, n^2 / \kappa. \label{eq:eH}
\een
The exchange energy of the uniform gas was derived in Ref.~\onlinecite{ECPG14}, with the energy per length as
\begin{equation}\label{energyex}
e\unpol\x(n)=A\kappa\left(\ln(1+y^2)-2y\,\mathrm{arctan}\,y\right)/(2\pi^2)
\end{equation}
where $y=\pi/(2\kappa r_s)$ where $r_s$ is the Wigner-Seitz radius \cite{AM05} in 1d defined as $r_s=1/(2n)$ \cite{GV08}.   
The limits for this expression are
\begin{eqnarray}\label{exlim}
e\x\unpol(n)\rightarrow&-A n/2,\quad& n\rightarrow\infty\\
e\x\unpol(n)\rightarrow&-A n^2/(2\kappa),\quad& n\rightarrow0.\label{exlimhigh}
\end{eqnarray}
Both the exchange and kinetic energies depend separately on the orbitals of each spin, so they satisfy simple spin scaling relation \cite{PK03}:
\ben
E\x[ n, \zeta]=\frac12\left\{E\unpol\x[(1+\zeta)n]+E\unpol\x[(1-\zeta)n]\right\}
\een
where $\zeta(x)=(n\up(x)-n\dn(x))/n(x)$ is the local spin polarization for spin up (down) densities $n\up$ ($n\dn$).  Applying the relation for $T\s$ yields:
\ben\label{tszeta}
t\unif\s(n,\zeta)=t_s\unpol(n)(1+3\zeta^2)/2
\een
and the exchange energy per length:
\ben\label{exzeta}
e\unif\x(n,\zeta)=\sum_{\sigma=\pm1}e\unpol\x\Big((1+\sigma\zeta)n\Big)/2.
\een

\begin{figure}
\includegraphics[width=\columnwidth]{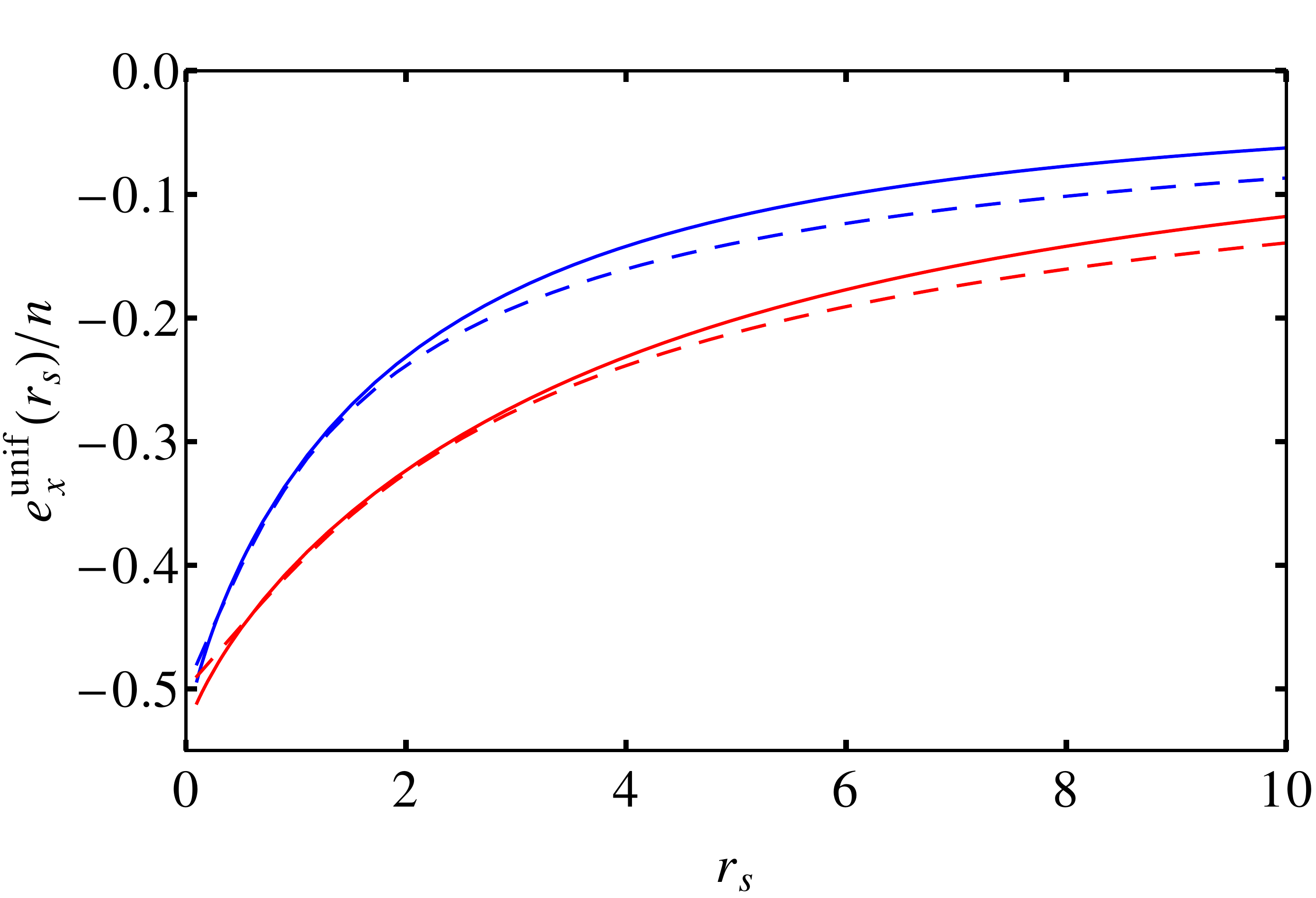}
\caption{(color online) Exchange energy densities per particle of the uniform gas for the exponential (solid lines), both unpolarized (blue) and fully polarized (red). Dashed lines show soft-Coulomb results.
\label{ex_combo}}
\end{figure}

In Fig.~\ref{ex_combo}, we plot the exchange energies per unit length for both the unpolarized and 
fully polarized cases, comparing with the soft-Coulomb.  For $r\s\lesssim2$, {\it i.e.} $n\gtrsim1/4$, they are very similar, but the exponential vanishes more rapidly in the low density limit. Applying Eq.~\eqref{exlimhigh} to Eq.~\eqref{exzeta} shows $e\x\unif$ is independent of $\zeta$ in the high-density limit.  Application of Eq.~\eqref{exlim} shows $e\x\unif(n,\zeta)\rightarrow e\x\unpol(n)(1+\zeta^2)$ as $n\rightarrow0$.

\subsection{Correlation Energy}

While we can determine the exchange energy of the uniform gas analytically, for correlation we cannot.  We first determine the high and low density limits and connect them with a Pad\'e approximation.

\subsection{High Density Limit}

In the high-density uniform gas, the RPA solution becomes exact \cite{NP58,GB57,LSJ92,SH96,LS91,FW71,GV05, FW71, Z72,LP77,RK77}.
Casula, {\it et.~al.} \cite{CSS06} have derived a criteria for the RPA
 limit which for 1d systems gives $e\c=-\Upsilon r\s/\pi^4$ where $\Upsilon=\int_0^\infty dq\,q\tilde{v}\ee^2(q)$ for the Fourier transform, $\tilde{v}\ee(q)=2A\kappa/(\kappa^2+q^2)$ \cite{CSS06,SCSM08,SCSM09,HFCV11}.
Thus,
\begin{equation}
e\c^\mathrm{RPA}(n)=\begin{cases}
-2A^2r_s/\pi^4&\mathrm{unpolarized},\\
-A^2r_s/(4\pi^4)&\mathrm{polarized}.
\end{cases}
\end{equation}
Note that these expressions for the exponential $e^\mathrm{RPA}\c$ are identical to the $e^\mathrm{RPA}\c$ of the soft-Coulomb interaction if $A=1$ and so differ by less than 15\% here \cite{HFCV11}.

\subsection{Low Density Limit}\label{lowdens}

To determine the low density limit of the correlation, we may work in the limit of a Wigner crystal \cite{W32}.  This phase occurs because the kinetic energy becomes less than the interaction energy for low densities. The strictly correlated electron limit \cite{SPL99,SGS07,MB04} is the exact limit of the low density electron gas \cite{SPK00b,GV08}.
For strictly correlated electrons, each electron sits some multiple of $2 r\s$
away from any other. For short-ranged interactions, such as the exponential,
the potential energy must decay exponentially with $r\s$, 
so that $E\xc = -U$, regardless of spin polarization \cite{PK03}. The actual form of
the correlation energy $e\c$, however, will depend on the polarization,
because the exchange energy $e\x$ does.  For a spin-unpolarized system,
using Eq.~\eqref{exlimhigh}, we find $e\x(n) = e\c(n)=-A n^2/(2\kappa)$ in the low-density gas limit.
After spin-scaling exchange, we obtain the low-density limit $e\x\pol(n)=-A n^2/\kappa $,
and this quantity already cancels the Hartree energy!
Therefore the correlation energy for spin-polarized electrons must decay more rapidly than $n^2$ as $n\rightarrow0$.

\subsection{Correlation Energy Calculations}

\begin{figure}
\includegraphics[width=\columnwidth]{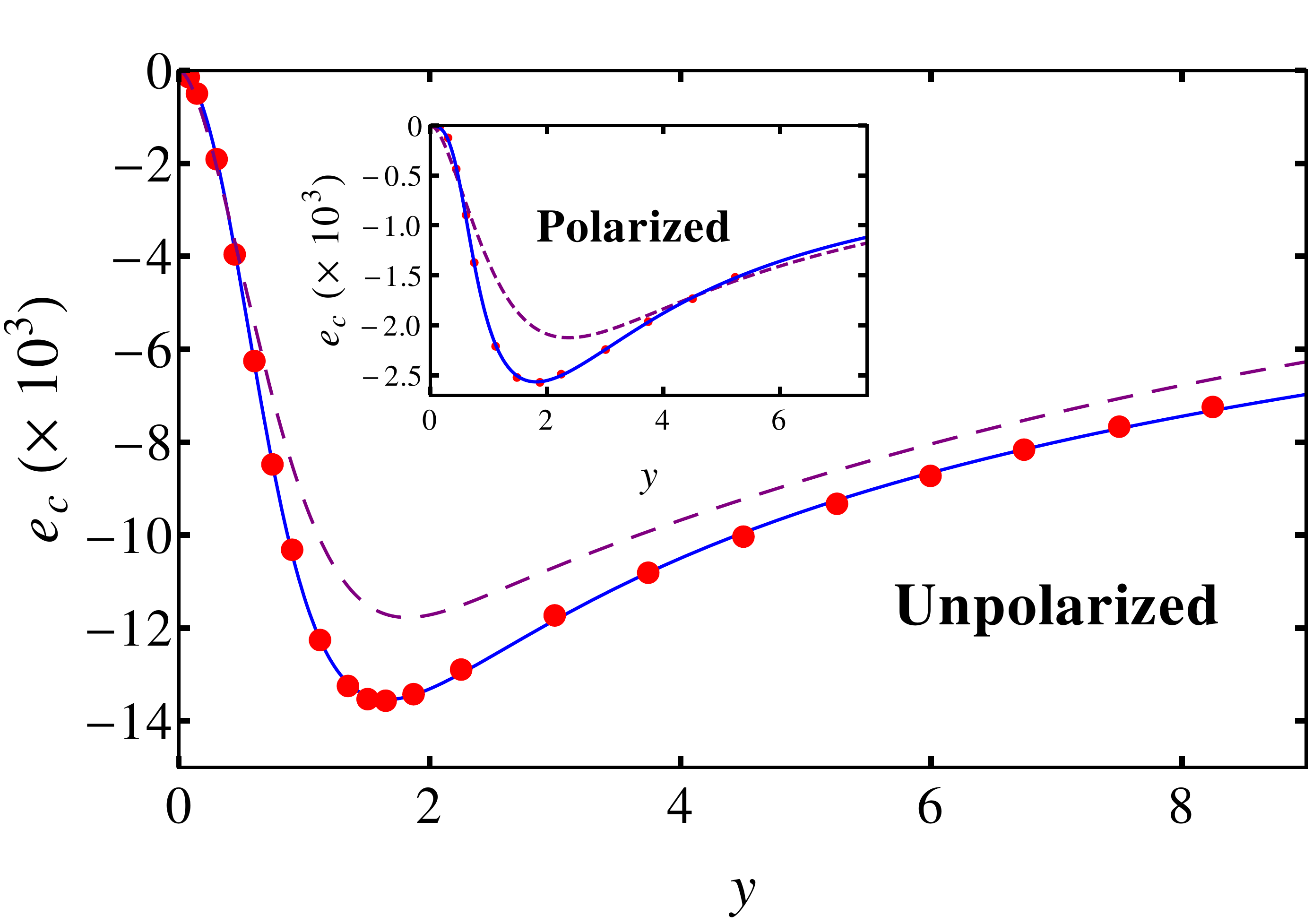}
\caption{(color online) 
Correlation energy densities of the uniform gas from DMRG calculations.  
Solid lines are parametrization of Eq.~\eqref{ecfit}, red dots are calculated, and
dashed lines are for the soft-Coulomb \cite{HFCV11}.
\label{fig: unpolfit}
}
\end{figure}

The uniform gas correlation energy was found from a sequence of DMRG calculations in boxes of increasing length, $L$, in which the average density is kept fixed.  Grid errors become more pronounced in these systems compared with soft-Coulomb interactions because there is no way to ensure a grid point will always lie on the wave function's cusp.  Hence, a very fine grid spacing was necessary to converge the points.  As the density increases, finer and finer grids were necessary, down to a practical limit of $\Delta=0.008$.
Energies for different box sizes were fit to a parabola in $1/L$ and
the limit of $L\rightarrow\infty$ extracted.  
The various energy components of Sec.~\ref{tshxc} were then subtracted to
find $e\c$ to make the dots in Fig.~\ref{fig: unpolfit}. 
For partially polarized gases, we used the same procedure as for the unpolarized case except that each box contained various values of $\zeta$ for each $L$ used in the limit.  Evaluating several partially polarized systems, we found that the correlation energy was very nearly parabolic in $\zeta$ at several different densities.

\subsection{Pad\'e Approximation}

Considering the asymptotic limits of the correlation energy,
the full approximation that allows us to accurately fit the data is
\ben\label{ecfit}
e\unif\c=\frac{-A\kappa y^2/\pi^2 }{\alpha+\beta \sqrt y+\gamma y+\delta y^{3/2}+\eta y^2+\sigma y^{5/2}+\n\frac{\pi\kappa^2}Ay^3}.
\een
where $y$ was defined above and the fitting parameters are defined for this specific choice of $A$ and $\kappa$ only. The parameters optimized by fitting
DMRG uniform gas data are given in Table~\ref{tab: Ecparams}, and the resulting fits are
shown in Fig.~\ref{fig: unpolfit}.  Although we know at full polarization $e\c$ should decay more rapidly than $n^2$ in the low density limit, we do not know how much more rapidly, so we use the same form as the unpolarized case.  This fit is accurate and the coefficient of $n^2$ is about 1\% that of the unpolarized one. Finally, we approximate the $\zeta$ dependence as
\begin{equation}\label{energyec}
e_c(n,\zeta)=e_c\unpol(n)+\zeta^2\Big(e_c\pol(n)-e_c\unpol(n)\Big),
\end{equation}
as justified in the previous section.

\begin{table}[htb]
\begin{center}
\begin{tabular}{| c || c | c | c | c | c | c | c }
\hline
& $\alpha$ & $\beta$ & $\gamma$ & $\delta$ & $\eta$ & $\sigma$ \\
\hline
  $e\unpol_c$   & 2 & -1.00077 & 6.26099 & -11.9041 & 9.62614 & -1.48334 \\
 $e\pol_c$ & 180.891 & -541.124 & 651.615 & -356.504 & 88.0733 & -4.32708 \\
\hline
\end{tabular}\\

\caption{Fitting parameters for the correlation energy in Eq.~\eqref{ecfit}. The parameter $\n=1$ (unpolarized) or 8 (polarized).\label{tab: Ecparams}}
\end{center}
\end{table}

\begin{figure}
\includegraphics[width=\columnwidth]{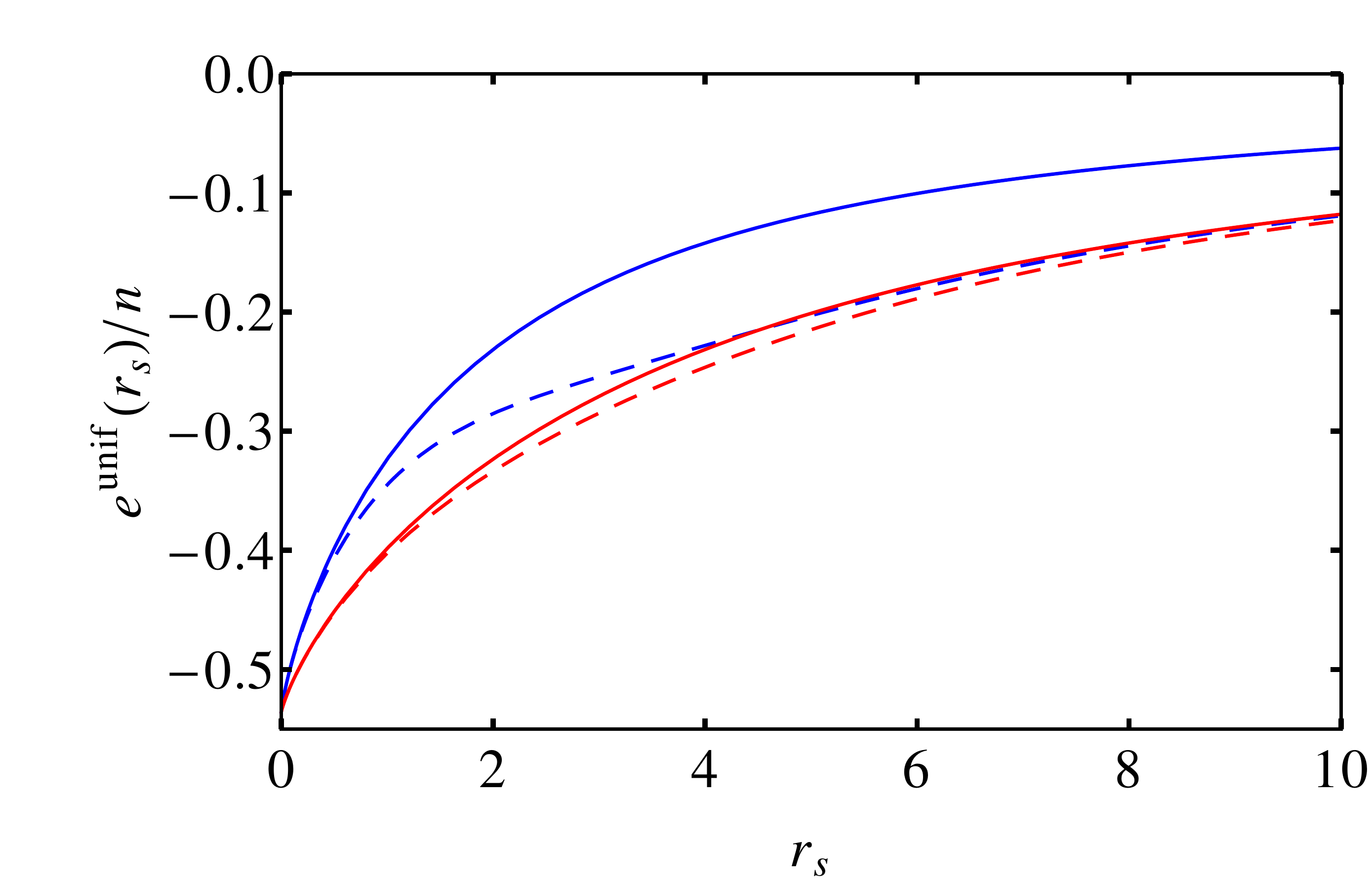}
\caption{(color online) 
XC energy per particle, $e\xc/n$ (dashed), for the unpolarized (blue) and fully
polarized (red).  Solid lines are $e\x/n$ only.
\label{fig: LDAXC}
}
\end{figure}
In Fig.~\ref{fig: LDAXC}, we show the combined
XC energies (solid) for the unpolarized (blue) and polarized gases (red).
Dashed lines are for exchange alone. Correlation is a much smaller effect in the fully spin-polarized
gas.

\section{Finite Systems}\label{results}

\subsection{Atomic Energies}\label{atoms}

\begin{table*}[htb]
\begin{minipage}{2\columnwidth}
\begin{center}
\begin{tabular}
{| l | l || c | c | c || c | c | c || c | c | c || c | c | c || c | c | c |}
\hline
$N\e$& Symb. &  $E^\mathrm{HF}$ & $E$ & $E\c^\mathrm{QC}$ & $T$ & $V$ & $V\ee$ & $T_s$ & $U$ & $E\xc$ & $E\x$ & $E\c$ & $T\c$ & $E\LDA$ & $E\LDA\x$ & $E\LDA\c$\\
\hline
1      & H         & -0.670 & -0.670 & 0 & 0.111 & -0.781 & 0 & 0.111 & 0.345 & -0.345 & -0.345 & 0 & 0 & -0.643 & -0.305 & -0.009 \\
       & He$^+$    & -1.482 & -1.482 & 0 & 0.190 & -1.672 & 0 & 0.190 & 0.379 & -0.379 & -0.379 & 0 & 0 & -1.449 & -0.337 & -0.007 \\
       & Li$^{++}$ & -2.334 & -2.334 & 0 & 0.259 & -2.593 & 0 & 0.259 & 0.397 & -0.397 & -0.397 & 0 & 0 & -2.298 & -0.355 & -0.006 \\
       & Be$^{3+}$ & -3.208 & -3.208 & 0 & 0.321 & -3.529 & 0 & 0.321 & 0.408 & -0.408 & -0.408 & 0 & 0 & -3.171 & -0.366 & -0.005 \\
\hline
2      & H$^-$       & -0.694 & -0.737 & -0.044 & 0.114 & -1.311 & 0.460 &  0.081 & 1.070 & -0.586 & -0.535 & -0.051 & 0.050 &  -0.711 & -0.520 & -0.073\\
       & He          & -2.223 & -2.237 & -0.014 & 0.286 & -3.212 & 0.690 &  0.273 & 1.432 & -0.730 & -0.716 & -0.014 & 0.013 &  -2.196 & -0.633 & -0.050\\
       & Li$^+$      & -3.884 & -3.892 & -0.008 & 0.433 & -5.080 & 0.755 &  0.426 & 1.541 & -0.779 & -0.770 & -0.009 & 0.007 &  -3.842 & -0.686 & -0.039\\
       & Be$^{++}$   & -5.606 & -5.611 & -0.005 & 0.564 & -6.967 & 0.792 &  0.559 & 1.602 & -0.806 & -0.801 & -0.005 & 0.005 &  -5.556 & -0.715 & -0.034\\
\hline
3      & Li       & -4.199 & -4.215 & -0.016 & 0.628 & -6.490 & 1.647 &  0.614  & 2.751 & -1.089 & -1.072 & -0.017 & 0.014 & -4.181 & -0.999 & -0.045\\
       & Be$^+$   & -6.447 & -6.457 & -0.010 & 0.910 & -9.225 & 1.858 &  0.900  & 3.029 & -1.161 & -1.150 & -0.011 & 0.010 & -6.411 & -1.074 & -0.035\\
\hline
4      & Be       & -6.756 & -6.809 & -0.053 & 1.118 & -11.115 & 3.188 &  1.077  & 4.710 & -1.481 & -1.421 & -0.060 & 0.041 & -6.784 & -1.371 & -0.080\\
\hline
\end{tabular}
\caption{Energy components for several systems as calculated by DMRG to 1 mHa accuracy. Chemical symbols H, He, Li, and Be imply single atomic potentials of $Z=1,2,3,$ and 4, respectively.  All LDA calculations are self-consistent, except H$^-$, which is unbound, so we evaluate $E\xc\LDA$ on the exact density \cite{KSB13}.  These values are very similar to those of the soft-Coulomb \cite{WSBW12}. \label{tab: manyeparams}}
\end{center}
\end{minipage}
\end{table*}

We now refer to several tables that contain the information from several
different systems as determined by the methods of the previous sections. 
To construct the LDA, 
$E\xc\LDA$ is constructed from Eqs.~\eqref{energyex} and \eqref{energyec}.
In practice, we always use spin-DFT, and the local spin density approximation (LSDA).
The plot of $e\unif\xc(r\s,\zeta)$
is very similar to those of the soft-Coulomb (see Fig.~2 of Ref.~\onlinecite{WSBW12}).

Table~\ref{tab: manyeparams} contains many energy components for all 1d atoms and ions up to $Z=4$.  The total energies are accurate to within 1 mHa.  The first approximation in quantum chemistry is the Hartree-Fock (HF), and its error is the (quantum chemical) correlation energy, $E\c^\mathrm{QC}$.  As required, this is always negative and is typically a very small fraction of $|E|$.  Unlike real atoms and ions, for fixed  particle number $N$, $E\c\rightarrow0$ as $Z\rightarrow\infty$, not a constant \cite{CGDP93}.

The next three columns show the breakdown of $E$ into its various components.  The magnitude of  $|V|$ is much greater than the other two.  Unlike 3d Coulomb systems, there is no virial theorem relating $E$ and $-T$, for example \cite{L08}.  Here the kinetic energies are much smaller than $|E|$, just as for the soft-Coulomb \cite{WSBW12}.
In the following three columns we give the KS components of the energy. These were extracted by finding the exact $v\s(x)$ from $n(x)$ \cite{WBSB14}, and constructing the exact KS quantities \cite{KS65}.  Unlike 3d Coulomb reality, $T\s$ is smaller in magnitude than $|E\xc|$.  But just like in 3d Coulomb reality, both $T\s$ and $E\xc$ always grow with $Z$ if $N$ is fixed, with $N$ if $Z$ is fixed, or with $Z$ for $Z=N$ \cite{B07}.

The next set of columns break $E\xc$ into its components: $E\x$, $E\c$, and $T\c$.  The correlation energy is negative everywhere and is slightly larger in magnitude than $|E\c^\mathrm{QC}|$ as required \cite{PG96}.  We also see $T\c$ is very close to $-E\c$ everywhere, a sign of weak correlation \cite{B06}, which is the same as in 3d Coulomb atoms and ions.
We also report self-consistent LDA energies and the XC components.  All LDA energies are insufficiently negative.  For $N=1$, self-interaction error occurs and $E\xc\LDA$ is not quite $-U$.  Just as for reality, LDA exchange consistently underestimates the magnitude of $E\x$, while $E\c\LDA$ overestimates, producing the well-known cancellation of errors in $E\xc\LDA$.

\begin{table}[htb]
\begin{center}
\begin{tabular}{|l | l || c | c || c | c || c |}
\hline
$N\e$& Symb.  & \multicolumn{2}{c||}{LDA}  & \multicolumn{2}{c||}{HF}   & Exact \\
\hline
 &             & $-\epsilon\H$ & I & $-\epsilon\H$ & I & I$=-\epsilon\H$ \\
\hline
1&  H           &  0.412  & 0.643 & 0.670 & 0.670 & 0.670  \\
\hline
2&  H$^-$       & -- & 0.062 & 0.058  &  0.024  & 0.068 \\
&  He          & 0.478 & 0.714  & 0.750  &  0.741 & 0.755   \\
&  Li$^+$      & 1.238 & 1.508  & 1.556  &  1.550  & 1.558   \\
&  Be$^{++}$   & 2.061 & 2.348  & 2.402  &  2.403  & 2.403   \\
\hline
3&  Li          & 0.182 & 0.339  & 0.327  & 0.315  & 0.323   \\
&  Be$^+$      & 0.643 & 0.855  & 0.850  & 0.836 & 0.846   \\
\hline
4&  Be          & 0.183 & 0.373 & 0.327 & 0.309 &  0.331  \\
\hline
\end{tabular}\\
\hspace{2cm}
\caption{Highest occupied eigenvalues $(\epsilon\H)$ and total energy differences ($I$) for several atoms and ion, both exactly and approximately.\label{tab: Igaps}}
\end{center}
\end{table}

We close the section on atoms with details of eigenvalues. It is long known that \cite{AvB85}, for the exact KS potential, the highest occupied eigenvalue is at $-I$, the ionization energy, but that this condition is violated by approximation.  In Table~\ref{tab: Igaps}, we list $-\epsilon\H$ and $I$ for several atoms and ions exactly, in HF and in LDA.  Koopman first argued \cite{K34} that $-\epsilon\H$ should be a good approximation to $I$ in HF, and we see that, just as in reality, it is a better approximation to $I$ than $I^\mathrm{HF}$, from total energy differences.  On the other hand, our 1d LDA exhibits the same well-known failure of real LDA: its KS potential is far too shallow, so that $\epsilon\H\LDA$ is above $\epsilon\H$ by a significant amount (up to 10 eV).

\subsection{Molecular Dissociation}

\begin{figure}
\includegraphics[width=1.1\columnwidth]{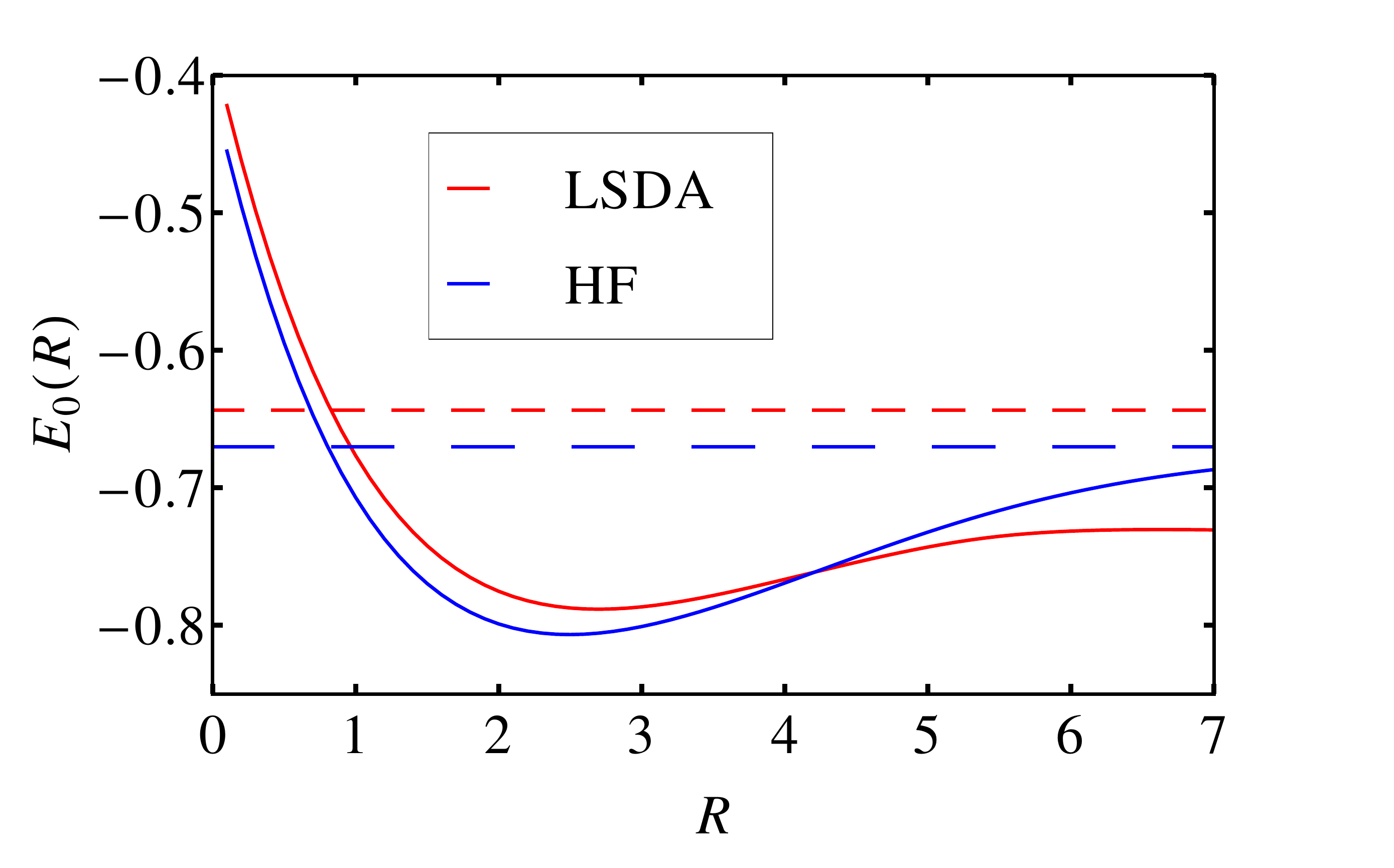}
\caption{(color online) Molecular dissociation for H$_2^+$ for LSDA and HF. Dashed lines indicate the energy of a single H atom.  Results are in close agreement with the soft-Coulomb interaction (see Fig.~1 of Ref.~\onlinecite{HFCV11} and Fig.~8 of Ref.~\onlinecite{WSBW12}).  LSDA has the well known failure of not dissociating to the correct limit \cite{B07}. \label{H2+}}
\end{figure}
\begin{figure}
\includegraphics[width=1.1\columnwidth]{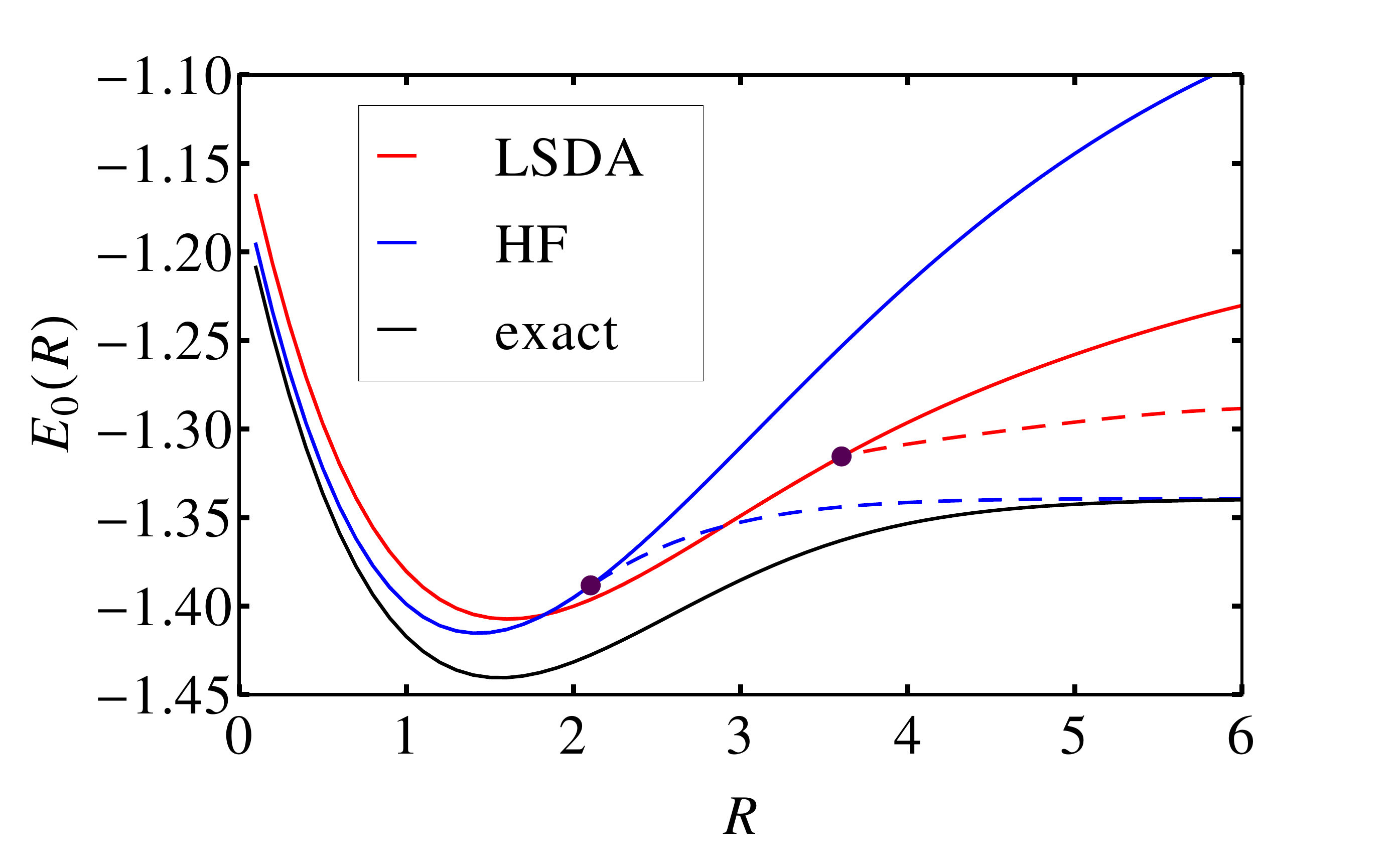}
\caption{(color online) Molecular dissociation for H$_2$ with dots denoting the Coulson-Fischer points.  Curves are shown for the exact case; also shown are restricted (solid) and unrestricted (dashed) LSDA and HF.\label{H2}}
\end{figure}

Next we consider 1d molecules with an exponential interaction.  The behavior is almost identical to that of the soft-Coulomb documented in Ref.~\onlinecite{WSBW12}.  For H$_2^+$, in Fig.~\ref{H2+}, HF is exact, and $E_0^\mathrm{LSDA}(R\rightarrow\infty)$ does not tend to $E_0^\mathrm{LSDA}(\mathrm{H})$ because of a large self-interaction error \cite{CMY08}. 

For H$_2$, in Fig.~\ref{H2}, the exact curve is calculated with DMRG, which has no problems whatsoever with stretching the bond (and can even break triple bonds \cite{CG02}).  But HF, restricted to a singlet, tends to the wrong limit as $R\rightarrow\infty$, while unrestricted HF goes to a lower energy beyond the Coulson-Fischer \cite{CF49} point ($R=2.1$) and dissociates to the correct energy $(2E(H))$ but the wrong spin symmetry \cite{PSB95}.  The same qualitative behavior occurs for LSDA at $R=3.6$.  These results are almost identical to those with soft-Coulomb interactions (except $R=3.4$ for the soft-Coulomb LSDA Coulson-Fischer point), and qualitatively the same as 3D.

For reference purposes, we also report equilibrium properties of H$_2$ and H$_2^+$ in Table~\ref{tab: H2nums}.  HF is exact for H$_2^+$, but underbinds H$_2$, shortens its bond, and yields too large a vibrational frequency, just as in 3d.  Overall, LSDA results are substantially more accurate.  LSDA overbinds, yields slightly (to less than 1\%) too large a bond, and only slightly overestimates vibrational frequencies, just as in reality.  Vibrational frequencies, $\omega$, are chosen to fit the function $E_0+\omega^2(r-R_0)^2/2$.

\begin{table}[htb]

\begin{tabular*}{\columnwidth}{@{\extracolsep{\fill}}|c|cc||ccc|}
\hline
&\multicolumn{2}{c||}{H$_2^+$}	& \multicolumn{3}{c|}{H$_2$}		\\
\hline
quantity & LSDA & exact & HF & LSDA & exact \\
\hline
$D\e$ (eV)                         & 3.94 & 3.72 & 2.04 & 3.25 & 2.74 \\
$R_0$ (bohr)                       & 2.70 & 2.50 & 1.45 & 1.60 & 1.56\\
$\omega$  ($\times10^3$ cm$^{-1}$) & 2.03 & 2.32 & 3.91 & 3.52 & 3.40 \\
\hline
\end{tabular*}

\caption{
Electronic well depth $D\e$ (calculated relative to well-separated unrestricted atoms), equilibrium bond radius $R_0$, and 
vibrational frequency $\omega$  for the H$_2^+$ and H$_2$ molecules.
\label{tab: H2nums}
}
\end{table}

\subsection{Relative Advantages}

\begin{figure}[htb]
\includegraphics[width=1.1\columnwidth]{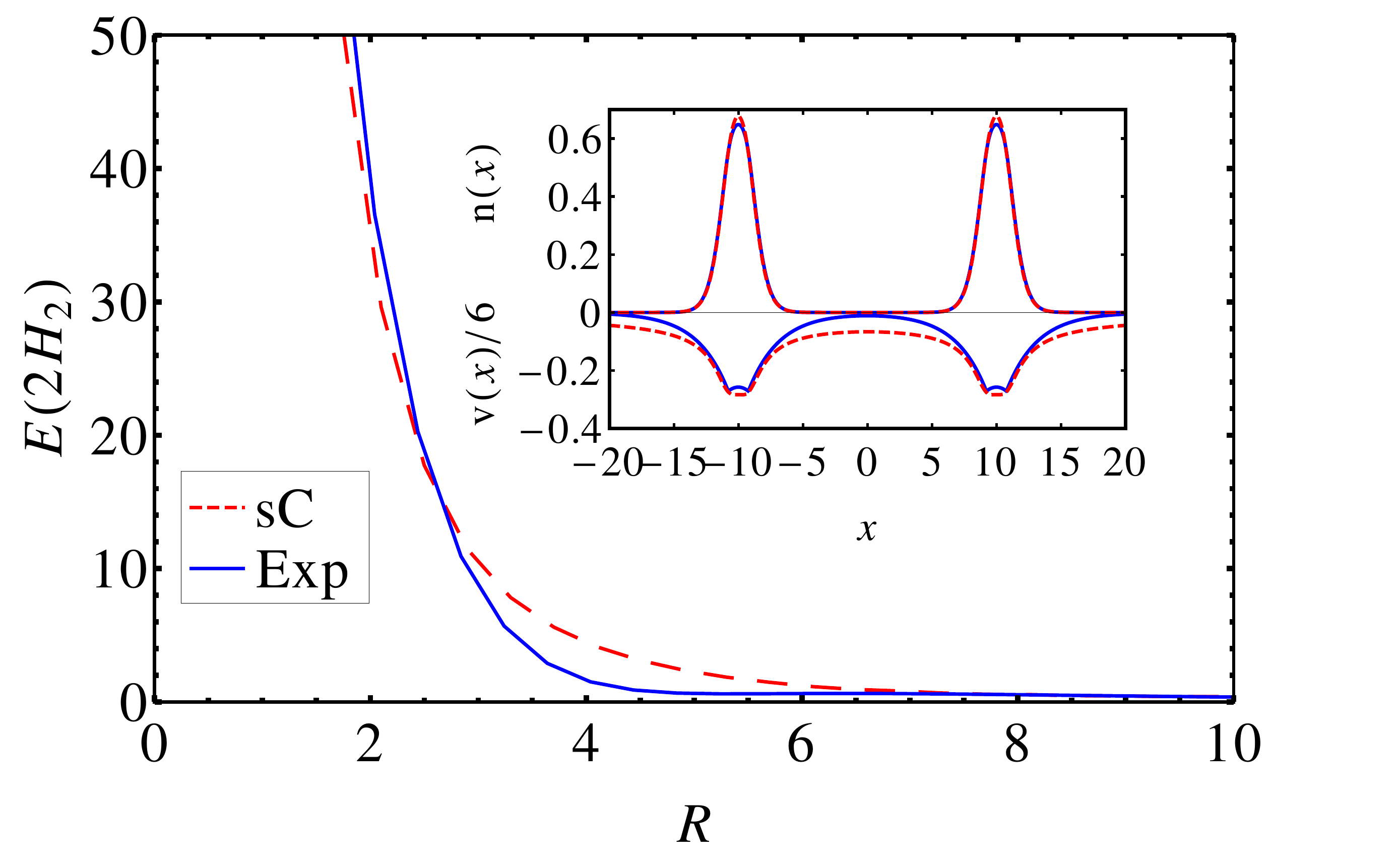}
\caption{(color online) 
Binding energy curves for two H$_2$ molecules
a distance $R$ apart (measured between the innermost atoms),
demonstrating the much slower decay of the soft-Coulomb. We subtracted off the asymptote of the binding curve so that the curves tend to zero. 
The inset shows the soft-Coulomb
potentials  from the two H$_2$ atoms have a substantial overlap even at $R=18$ while the exponential does not.   
\label{fig: EH2}
}
\end{figure}

To illustrate the effective shorter range of the exponential relative to 
a soft-Coulomb, Fig.~\ref{fig: EH2} shows the exact binding energy of
two H$_2$ molecules (with bond lengths set to their
equilibrium values for each type of interaction)
as a function of the separation between closest nuclei.
These closed shell molecules do not bind with either interaction, but
the soft-Coulomb energy decays far more slowly to its value at $\infty$.

\begin{figure}[htb]
\includegraphics[width=1.1\columnwidth]{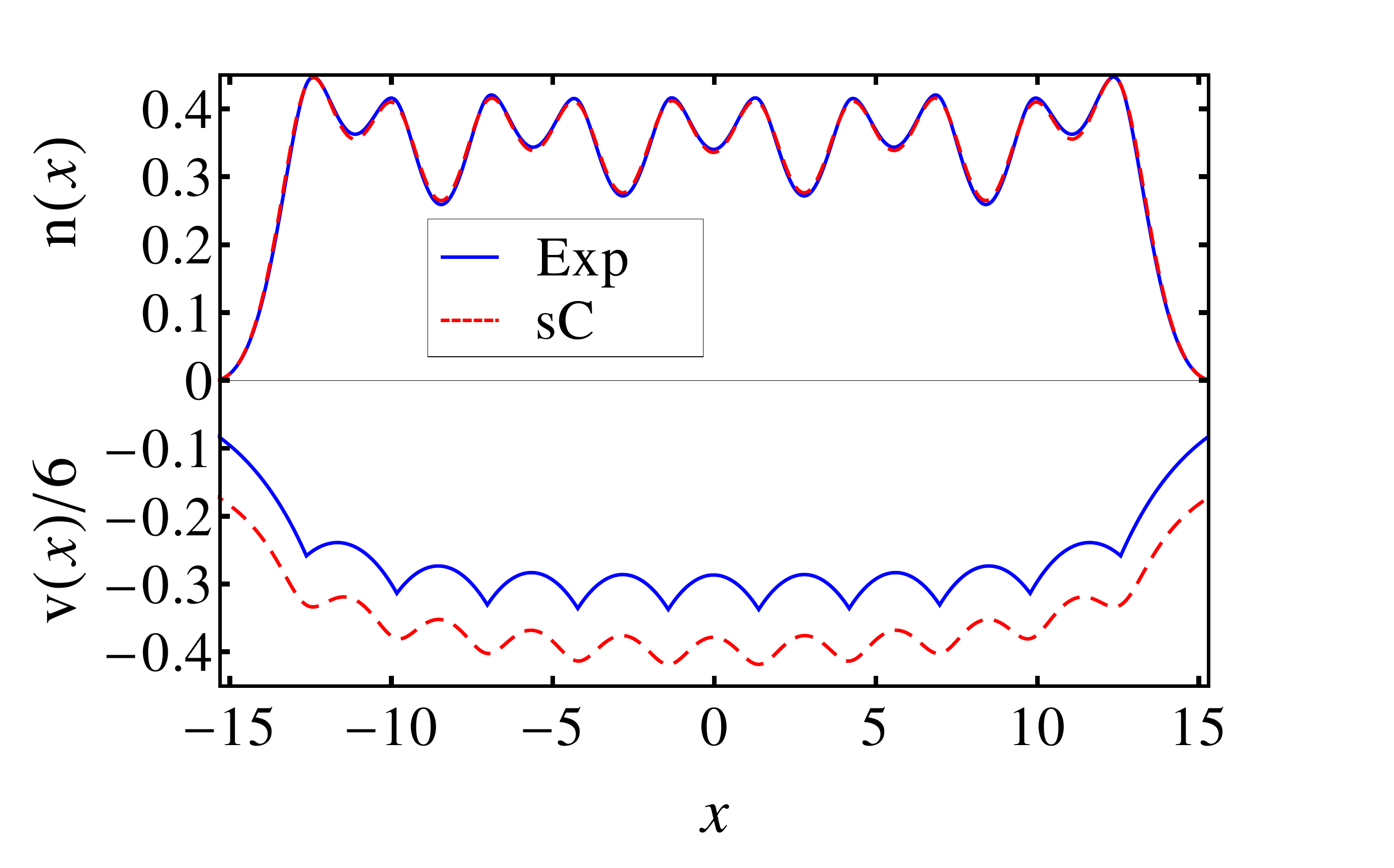}
\caption{(color online) 
The density of an H$_{10}$ chain with
$R=2.8$ with exponential and soft-Coulomb interactions.  
\label{fig:chain}
}
\end{figure}
\begin{figure}[htb]
\includegraphics[width=1.1\columnwidth]{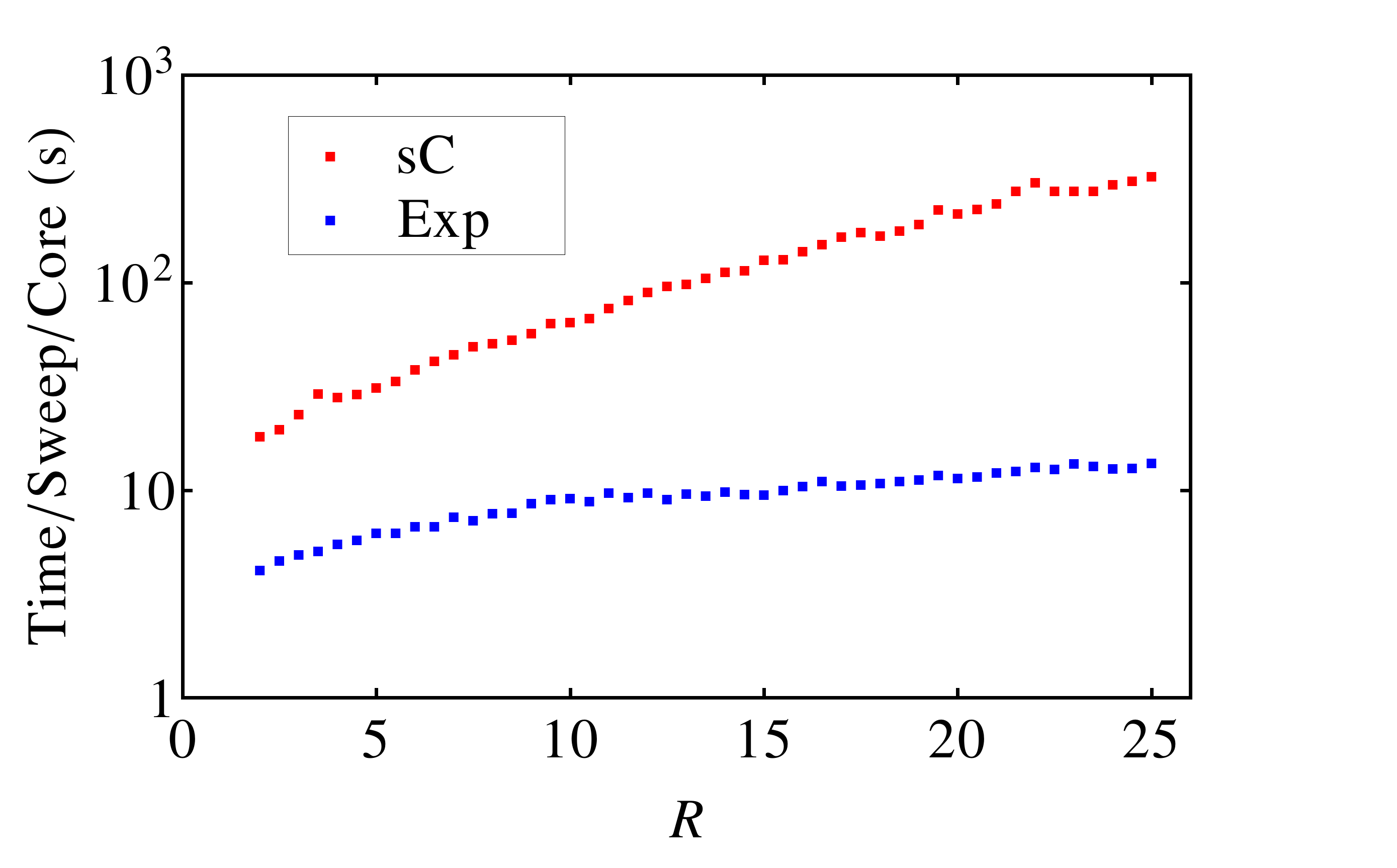}
\caption{(color online) 
The exponential performs sweeps in DMRG much faster, leading to a greatly reduced computational time.  Here we show the cost of the H$_{10}$ of Fig.~\ref{fig:chain} as a function of interatomic separation.  The number of many body states is fixed at 30 for 10 sweeps and each system has 1161 grid points.  The maximum truncation error in each sweep is $\sim10^{-8}$.
\label{fig: time}
}
\end{figure}

Finally, we give an example of the much lower
computational cost of the exponential in DMRG over that of
the soft-Coulomb due to the reasons discussed in Sec.~\ref{intro}.  
In Fig. \ref{fig:chain}, we plot the densities and potentials for
a 10-atom H chain with separations $R=2.8$.   The 
densities are sufficiently similar for all practical purposes
(but not the potentials) while the computational time per 
DMRG sweep was about 4 times longer for the soft-Coulomb. With increasing interatomic distance, Fig.~\ref{fig: time} shows a decreased wall time of more than a factor of 20 for calculating larger separations.

\section{Conclusion}

We have introduced an exponential interaction that allows us to mimic many features of real electronic structure with 1d systems.  Using the exponential in place of the more standard soft-Coulomb interaction not only improves the computational time of DMRG , but it also allows faster convergence for other calculations due to its fast decay and local nature.  To facilitate comparison with existing calculations, we choose the parameters in the exponential to best match a soft-Coulomb potential.  A parameterization for the LDA is given by calculating and fitting to uniform gas data.

\section{Acknowledgments}
This work was supported by the U.S. Department of Energy, Office of Science, Basic Energy Sciences under award \#DE-SC008696.  T.E.B.~also thanks the gracious support of the Pat Beckman Memorial Scholarship from the Orange County Chapter of the Achievement Rewards for College Scientists Foundation.

\bibliography{exp,Master,notes}

\begin{thebibliography}{99}
\bibitem [{exp()}]{expEnote}%
  \BibitemOpen
  \bibfield  {title} {\enquote {\bibinfo {title} {References~\onlinecite{ESJ89}
  and \onlinecite{WSBW12} gives the ground state energy as -0.669777, but a
  truncation in the variables $a$ and $\kappa$ change our energy slightly.}}\
  }\href@noop {} {\ }\BibitemShut {NoStop}%
\end{thebibliography}
\label{page:end}

\end{document}